\documentclass[preprint,12pt]{elsarticle}

\usepackage{amssymb}

\journal{---}

\begin{document}

\begin{frontmatter}



\title{A Class of Self-Trapped and Self-Focusing Wave Functions in Madelung Fluid Picture of A Single Free Particle Quantum System}


\author{Agung Budiyono and Ken Umeno}

\address{Institute for the Physical and Chemical Research (RIKEN)\\ 2-1 Hirosawa, Wako-shi, Saitama 351-0198, Japan}

\begin{abstract}

Using the Madelung fluid picture of Schr\"odinger equation for a single free particle moving in one spatial dimension, we shall specify a class of wave functions whose quantum probability density is being trapped by the quantum potential it itself generates. The global convexity of the quantum potential generated by the initial self-trapped wave function will then be shown to further localize the quantum probability density through a self-generated focusing equation for a finite interval of time. 

\end{abstract}


\end{frontmatter}


\section{Motivation: Schr\"odinger Equation as Self-Referential Dynamics\label{motivation}}

Let us consider a quantum system of a single free particle with mass $m$ moving in one dimensional space, $q$. The wave function $\psi(q;t)$, where $t$ is time, is then governed by the linear Schr\"odinger equation 
\begin{equation}
i\hbar\partial_t\psi(q;t)=-\frac{\hbar^2}{2m}\partial_q^2\psi(q;t).
\label{Schroedinger equation}
\end{equation}
Quantum theory claims that the quantum probability density, $\rho(q)=|\psi(q)|^2$, gives the essential information about the position of the particle \cite{interpretation,Isham book,Bohm-Hiley book,Bell unspeakable}. Hence, it is important to study the dynamics of $\rho(q;t)$. 

One way to keep track directly the dynamics of $\rho(q;t)$ is to use Madelung fluid hydrodynamical interpretation of the Schr\"odinger equation \cite{Madelung 1926}. To do this, let us write the wave function in polar form, $\psi(q;t)=R(q;t)\exp(iS(q;t)/\hbar)$, where the quantum amplitude $R$ and the quantum phase $S$ are real-valued functions. Inserting this into Eq. (\ref{Schroedinger equation}) and separating the real and imaginary parts, one gets
\begin{eqnarray}
\partial_tS+\frac{(\partial_qS)^2}{2m}+U=0,\nonumber\\
\partial_t\rho+\partial_q\Big(\frac{\partial_qS}{m}\rho\Big)=0,\hspace{2mm}
\label{pilot-wave dynamics}
\end{eqnarray}
where $U$ is the so-called quantum potential generated by the quantum probability density as 
\begin{equation}
U(q;t)=-\frac{\hbar^2}{2m}\frac{\partial_q^2R}{R}.
\label{quantum potential}
\end{equation}
Next, let us define a velocity field as
\begin{equation}
v(q;t)=\frac{1}{m}\partial_qS(q;t). 
\label{velocity field}
\end{equation}
Taking the spatial gradient to both sides of the upper equation in (\ref{pilot-wave dynamics}), and using Eq. (\ref{velocity field}) one has
\begin{equation}
m\frac{dv}{dt}=-\partial_qU.
\label{Newtonian equation}
\end{equation}
Due to its similarity with Euler equation, the term on the right hand side of the above equation, $-\partial_qU$, is usually called as quantum force field. Moreover, again using Eq. (\ref{velocity field}), the lower equation in (\ref{pilot-wave dynamics}) can be seen as the continuity equation for $\rho(q;t)$. 

One of the great lesson from the Madelung fluid picture is its self-referential property. Namely, the quantum probability density $\rho(q;t)$ will generate  a quantum potential $U(q;t)$ which in turn will rule the quantum probability density the way it must flow and so on and so forth. There is a dynamics circularity between the object to be ruled and the rule itself. Self-referential property is a simpton of complex non-linear systems, and is argued to be the general origin for the emergence of many interesting phenomena \cite{self-referential dynamics and complexity}. We can then expect to find interesting self-organized behavior difficult to be observed directly from Eq. (\ref{Schroedinger equation}), at least in a certain time scale. 

To see an example of the effect of the self-referential property of the Madelung fluid and to help describing the motivation of the present work, let us discuss the dynamics of a Gaussian wave packet. In this case the quantum probability density will take the form \cite{quantum mechanics book} 
\begin{equation}
\rho(q;t)=\frac{1}{Z}\exp\Big(-\frac{q^2}{2\sigma_t^2}\Big),\hspace{2mm}\sigma_t^2=\sigma^2+\frac{\hbar^2t^2}{4m^2\sigma^2}.
\label{Gaussian wave packet}
\end{equation}
Here, $Z$ is the trivial normalization constant, $\sigma_t$ is the width of the wave packet at time $t$ and $\sigma$ is its width at $t=0$. The corresponding quantum potential can thus be calculated to give 
\begin{equation}
U(q;t)=-\frac{\hbar^2 q^2}{8m\sigma_t^2}+\frac{\hbar^2}{4m\sigma_t}. 
\label{quantum potential for Gaussian wave packet}
\end{equation}

One can clearly see from the global concavity of the quantum potential given above that the quantum force field is repelling the quantum probability density to spread out. As time goes, the quantum probability density keeps its Gaussian form but is rescaled to be wider. This will also rescale the quantum potential to be wider while keeping its form, thus keeps the effect of repelling the quantum probability density. This simple picture is missing in the original Schr\"odinger equation of (\ref{Schroedinger equation}). Hence, Gaussian wave packet provides an example of a class of wave functions being repelled by its own self-generated quantum potential. 

The question that we want to discuss in this paper is then: \textit{is there any class of wave functions whose quantum probability density is self-trapped or self-attracted by the quantum potential it itself generates?} 

\section{Self-Trapped quantum probability density}

Obviously, the self-trapped quantum probability density that we want to develop must decay as the self-generated quantum potential increases. One therefore has to develop another relation between $\rho$ and $U$ which provides the supplement to Eq. (\ref{quantum potential}). There are many possibilities to create an $\rho-U$ relation which satisfies the above mentioned condition. In this paper, we shall consider a class of wave functions $\{\psi\}$, whose quantum probability density is given as the exponential of its quantum potential as follows
\begin{equation}
\rho(q;t)=\frac{1}{Z(\beta)}\exp\big(-\beta U(q;t)\big), 
\label{canonical quantum probability density}  
\end{equation}
where $\beta$ is a positive real-valued contant, and $Z$ is a normalization factor given by $Z=\int dq \exp(-\beta U)$. It is then clear that $\rho$ decays as $U$ grows. Notice that Eq. (\ref{canonical quantum probability density}) together with the definition of quantum potential given in Eq. (\ref{quantum potential}) comprise a differential equation for $\rho(q)$ or $U(q)$, subjected to the condition that $\rho(q)$ must be normalized.

In term of $U$, taking the spatial derivative twice to both sides of Eq. (\ref{canonical quantum probability density}) and recalling the definition of quantum potential given in Eq. (\ref{quantum potential}) one obtains 
\begin{equation}
\partial_q^2U=\frac{\beta}{2}(\partial_qU)^2+\Lambda^2U,
\label{NPDE for quantum potential}
\end{equation}
where $\Lambda=\sqrt{4m/\hbar^2\beta}$. First, one can see from the above equation that the positivity of quantum potential, $U>0$, guarantees its local convexity, $\partial_q^2U>0$. Further, it is clear that the above differential equation is invariant under the transformation $q\rightarrow -q$. These facts will guide us later to find a class of solutions which are positive everywhere (thus convex everywhere), and symmetric with respect to the vertical line $q=0$. 

\begin{figure}[tbp]
\begin{center}
\includegraphics*[width=6cm]{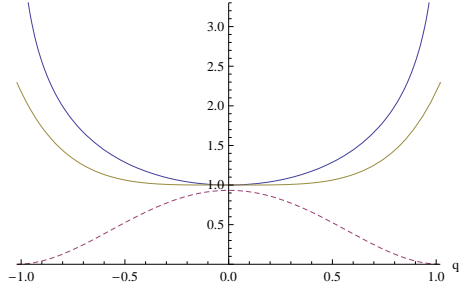}
\end{center}
\caption{The profile of quantum potential $U(q)$ (solid blue line) satisfying Eq. (\ref{NPDE for quantum potential}), and the corresponding quantum probability density $\rho(q)$ (dashed line) which generates the quantum potential. The solid red curve is the approximation of $U(q)$ given by Eq. (\ref{near extremum quantum potential}).}
\label{self-trapped quantum probability density}
\end{figure}

Let us first solve Eq. (\ref{NPDE for quantum potential}) for a spatial region very close to the extremum points of the quantum probability density or quantum potential, $|q-Q|\ll 1$, where $Q$ is the extremum point under consideration satisfying $\partial_qU(Q)=\partial_qU|_Q=0$. In this case, assuming the continuity of $\partial_qU$, the first term on the right hand side can be ignored, $\partial_qU\approx 0$. Hence, in this spatial region we have
\begin{equation}
\partial_q^2U\approx \Lambda^2U, \hspace{5mm}|q-Q|\ll 1.
\label{NPDE for quantum potential near extremum points}
\end{equation}
The above differential equation can then be solved with constraint $\partial_qU|_Q=0$ and $U(Q)=U_Q$, to give
\begin{equation}
U(q)\approx U_Q\cosh \big(\pm\Lambda(q-Q)\big), \hspace{5mm}|q-Q|\ll 1. 
\label{near extremum quantum potential}
\end{equation}
The point $Q$ turns out to be the local minimum (maximum) point of the quantum potential $U(q)$ if $U_Q$ is positive (negative). 

To check the desirable trapping property, one must see the global behavior of the quantum probability density and its corresponding quantum potential. To do this one should solve equation Eq. (\ref{NPDE for quantum potential}) for the whole space. We shall resort to numerical methods to do this. For the reason mentioned previously, we shall confine ourselves to a class of solutions which satisfy the following boundary conditions $\partial_qU|_0=0$ and $U_0>0$. Namely, we want to find a class of solutions where the first spatial derivative at the origin is vanishing, and let $U_0$ to varies only to positive values. The first condition will explore the symmetry of the differential equation, while the second boundary condition will guarantee that the quantum potential is nowhere non-convex. Note that the case $U_0=0$ is trivial. Namely, imposing the boundary condition $\partial_qU|_0=0$ to Eq. (\ref{NPDE for quantum potential}) one gets $\partial_q^2U|_{0}=0$ such that $U(q)=0$ everywhere. Hence, $\rho(q)$ is unnormalizable. This case therefore is excluded from our consideration. Fig. \ref{self-trapped quantum probability density} shows the numerical solution of Eq. (\ref{NPDE for quantum potential}) with the boundary condition $\partial_qU|_0=0$ and $U_0=1$. The profile of quantum probability density (dashed line) is plotted together with the profile of quantum potential it itself generates (solid blue line). For comparation, we have also plotted the approximate solution given in Eq. (\ref{near extremum quantum potential}) (solid red line).

First, it is observed that very near to the extremum point, $q=0$, Eq. (\ref{near extremum quantum potential})  fits the numerical solution very well. Yet it immediately departs from the numerical solution as $|q|$ is getting larger. We can see clearly from the numerical solution that globally the quantum probability density is indeed being trapped by the quantum potential it itself generates. This global property can be understood for any positive value of $U_0> 0$ as follows. First from Eq. (\ref{NPDE for quantum potential}), inserting the boundary conditions at $q=0$, one gets $\partial_q^2U|_0=\Lambda U_0 > 0$, such that $U(q)$ is locally convex at $q=0$. Hence, since $\partial_qU|_0=0$, at spatial points nearby $q=0$ one has $U(q)>U_0>0$. Moreover since the second term on the right hand side of Eq. (\ref{NPDE for quantum potential}) is always non-negative, at this region one has $\partial_q^2U>0$. This geometrical reasoning can be extended such that for the whole space one gets
\begin{equation}
\partial_q^2U>0. 
\label{convex quantum potential}
\end{equation}
Namely, the case when $U_0 > 0$ will give an everywhere positive and convex quantum potential. Hence, $q=0$ turns out to be the global minimum of $U(q)$. Using this fact in Eq. (\ref{canonical quantum probability density}) one can conclude that quantum probability density, $\rho(q)$, is being trapped by its own self-generated quantum potential, $U(q)$. 

Next, from the definite positivity of the quantum potential and Eq. (\ref{quantum potential}), one has
\begin{equation}
\partial_q^2R<0.
\label{concave quantum amplitude}
\end{equation}
Hence, the quantum amplitude is everywhere concave. Supplied with the fact that $R(q)$ is finite, one can conclude that $R(q)$ must cross the $q-$axis at finite value of $q=\pm q_m$. This means that the support of $R(q)$, $[-q_m,q_m]$ is finite. At $q=\pm q_m$, $U(q)$ is infinite, yet the quantum probability density is vanishing. Further, notice that at $q=\pm q_m$, $\partial_qR(q)$ is discontinuous. However, $\partial_q\rho(q)=2R(q)\partial_qR(q)=0$ is continuous. Hence, the self-trapped wave functions is a truly localized wave function with finite tail (support). 

\begin{figure}[tbp]
\begin{center}
\includegraphics*[width=6cm]{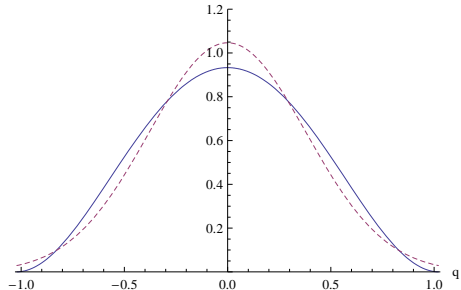}
\end{center}
\caption{The profile of a self-trapped quantum probability density (solid line) compared with a Gaussian quantum probability density (dashed line) with the same second moment.}
\label{self-trapped vs Gaussian}
\end{figure}

Now let us compare the class of self-trapped wave functions we developed in this paper with Gaussian wave packet. As discussed in section \ref{motivation}, the latter represents a class of self-repelled wave functions. Fig. \ref{self-trapped vs Gaussian} shows a self-trapped quantum probability density (solid line) plotted together with the Gaussian wave packet (dashed line) of the same second moment. One can see that Gaussian is more localized in the region near the maximum. Yet, in contrast to the self-trapped quantum probability density which possesses support only on a finite region, the Gaussian wave packet possesses support on the whole space. 

Another interesting difference between Gaussian and self-trapped wave function is that though both are quite similar to each other, their corresponding quantum potentials are very contrast from each other. While the quantum potential generated by Gaussian wave packet is concave everywhere, the quantum potential of self-trapped wave function developed in this paper is convex everywhere. As will be discussed in the next section, one can then expect that this makes their future destiny very different from each other at least for a finite period of time.  

\section{Self-Generated Focusing Equation}

Intuitively, for spatially one dimensional case we are considering, the global convexity of the quantum potential makes the quantum force field to drag the quantum probability density toward the center, \textit{i.e.} the global minimum of the quantum potential, at least for some interval of time. To see this more clearly, let us take the spatial derivative to both sides of Eq. (\ref{Newtonian equation}) to give us
\begin{equation}
md_t\theta=-m\theta^2-\partial_q^2U.
\label{focusing equation}
\end{equation} 
Here $d_t\equiv d/dt$, $\theta=\partial_q v=\partial_q^2S/m$ is the velocity divergence and we have used the following fact from the differential calculus $d_t=\partial_t+v\partial_q$. Hence, for the class of self-trapped wave functions we are considering here, where $\partial_q^2U>0$ everywhere, one will get 
\begin{equation}
d_t\theta <0,
\label{focusing equation}
\end{equation}
which is a focusing equation. Let us note that the above relation is valid for any value of $\theta(q)$, hence for any form of quantum phase $S(q)$. One can thus conclude that in contrast to Gaussian wave packet which is self-repelled everywhere, the self-trapped wave function is self-focused everywhere. 

Due to the continuity of the evolution of $U(q;t)$ which is guaranteed by the continuity equation in the lower part of (\ref{pilot-wave dynamics}), $U(q;t)$ will keep its global convexity for a certain finite interval of time, say $T$. Given an initial self-trapped quantum probability density, $\rho(q;0)$, $T$ of course depends on the initial quantum phase as well. Thus it depends on $\theta_0$, $T=T(\theta_0)$, where $\theta_0$ is the value of $\theta=(1/m)\partial_q^2S$ at $t=0$. Yet, the finite-ness of $T$ is independent of the choice of $\theta_0$. Namely, no initial wave function can avoid the focusing equation of (\ref{focusing equation}). In particular, during the interval of time $T$ one has
\begin{equation}
d_t\theta+\theta^2=-(1/m)\partial_q^2U \leq 0.
\label{caustics 1}
\end{equation}   
This will imply $d_t(\theta^{-1}) \geq 1$, such that one obtains
\begin{equation}
\theta^{-1}(t) \geq \theta_0^{-1}+t.
\label{caustics 2}
\end{equation}
The above inequality shows that if $\theta_0$ is negative, then  within a time $t_c < 1/|\theta_0|$, $\theta^{-1}$ must pass zero, namely, one has $\theta\rightarrow -\infty$. Of course this will occur as long as 
\begin{equation}
1/\theta_0 \leq T(\theta_0),
\label{inequality for self-collapsing}
\end{equation}
posseses a solution. Hence, caustics will develop in a finite interval of time. During this time, the quantum probability density keeps converging toward the origin.

\section{Conclusion and Discussion}

To conclude, first we have specified a class of wave functions whose quantum probability densities are being trapped by the quantum potentials they themselves generate. The self-trapped wave functions are then dragged by a self-generated focusing equation of (\ref{focusing equation}) to localize even further, at least for certain interval of time depending on the initial quantum phase. 

For a given initial self-trapped quantum probability density, if there is $\theta_0$ such that inequality (\ref{inequality for self-collapsing}) is satisfied, then the velocity divergence is equal to minus infinity everywhere, $\theta\rightarrow -\infty$, in a finite time. During this period of time, the quantum probability density converges everywhere until caustics is developed. The next question is then to investigate whether the singularity in velocity field (caustic) will also induce a singularity in quantum potential and how the quantum probability density looks like at this time. It is also tempting to study what happens to the wave function after the development of caustics. We leave this for future work.


\begin{thebibliography}{00}
\bibitem{interpretation}In the so-called orthodox interpretation of quantum mechanics, $\rho(q)$ is given meaning as the probability that a particle will be found at position $q$ if a measurement is performed. Where as in the ontological pilot-wave interpretation, $\rho(q)$ is interpreted as the probability that the particle is actually at $q$ regardless of any measurement. See for example \cite{Isham book,Bohm-Hiley book,Bell unspeakable}.
\end{thebibliography}
\end{document}